# Hot Carrier Dynamics in InAs-AlAsSb Core-Shell Nanowires

*Daniel Sandner[1], Hamidreza Esmaielpour[2], Fabio del Giudice[2], Matthias Nuber[1], Reinhard Kienberger[1], Gregor Koblmüller[2,*] and Hristo Iglev[1,*]*

[1] Chair for Laser and X-ray Physics, Physics Department, Technical University of Munich, James-Franck-Str. 1, 85748 Garching, Germany

[2] Walter Schottky Institute and Physics Department, Technical University of Munich, Am Coloumbwall 4, 85748 Garching, Germany

- Corresponding authors: E-Mail: hristo.iglev@tum.de; gregor.koblmueller@wsi.tum.de

# Abstract

Semiconductor nanowires (NWs) have shown evidence of robust hot carrier effects due to their small dimensions. The study of the dynamics of high-energy carriers, generated by photo-absorption, is of significant importance in optoelectronic devices and high efficiency solar cells, such as hot carrier solar cells. Among various III-V semiconductors, indium arsenide (InAs) NWs are promising candidates for their applications in advanced light harvesting devices due to their high photo-absorptivity and high mobility. Here, we investigate the dynamics of hot carriers using ultrafast pump-probe spectroscopy with widely tuned pump and probe energies. Through these measurements, we have studied the dynamics of photo-generated carriers and their interactions with longitudinal optical (LO) phonons in InAs-AlAsSb core-shell NWs, as well as bare-core InAs NWs. In addition, we have studied the electronic states in the AlAsSb-shell and found that despite the large band offset of the core-shell design in the conduction band, excited carriers in the shell remain there, leading to relatively long recombination lifetimes (>100 ps). Our results indicate evidence of plasmon-tailored core-shell NWs for efficient light harvesting devices, which can open up potential avenues for improving the efficiency of photovoltaic solar cells.

In recent years, advances in photovoltaic solar cells have led to the fabrication of single-junction devices with power conversion efficiencies close to the Shockley-Queisser limit (33%) under one sun condition. To further improve efficiency beyond this fundamental limit, major loss mechanisms in solar cells must be inhibited[1]. Although the thermalization of photo-generated hot carriers is considered one of the main loss mechanisms, it is still one of the least understood phenomena in semiconductors. Thermalization rates of hot carriers in nanostructured materials, such as NWs, are slowed due to locally confined charged carriers in these designs[2–4]. A comprehensive understanding of the origin of robust hot carriers and their interactions with phonons in these nanostructures is of significant importance for the design of efficient hot carrier absorbers with limited thermalization loss.

Photoexcited carriers in semiconductor nanostructures have been studied intensely for a variety of applications in light harvesting and emission[5–7]. For light emitters and detectors, key factors determining their efficiency are the lifetime and recombination mechanisms of charge carriers, which are readily studied by time resolved spectroscopy and a variety of other luminescence experiments, as shown for diverse nanostructured materials, including NWs[8,9]. Remarkable increases in the radiative recombination efficiency, carrier lifetime and mobility were reported in NWs over the years, when their large surface trap densities are passivated in the form of core-shell NW structures[8,10,11]. Overcoating NWs with such shell layers also introduces very versatile means to tailor emission properties via strain and the nature of band alignment[12–14]. Apart from carrier recombination, cooling of hot carriers is of great interest for novel solar cell concepts which seek to overcome the Shockley-Queisser limit[15]. Absorber materials can be either tuned to a low bandgap, thus, most solar photons can be absorbed, or to a large bandgap which allows utilization of high energy photons. If cooling of hot carriers can be slowed down to timescales where extraction is feasible, excess energy can be used and large efficiencies for light conversion are expected[16]. Since phonons play a major role in cooling of electrons, great effort has been invested into investigating phonon bottlenecks in nanocrystals and quantum wells[4]. Two kinds of phonon bottlenecks turned out to be possible: Hot phonon bottlenecks that rely on isolated LO-phonon populations, and which are in thermal equilibrium with hot electrons if a certain threshold is reached[17]. Phonon bottlenecks due to quantization overcome the cooling via phonons entirely *via* strong quantization of the conduction band (CB) – thus, electronic states are separated by multiple phonon energies[18]. However, hot electrons can circumvent these bottlenecks *via* other cooling channels, particularly by transferring energy to holes which have enhanced phonon scattering rates.[19,20].

Amongst the variety of candidate materials hosting such prominent hot carrier and phononic effects, narrow-band gap semiconductors with a large difference in electron and hole effective masses ($m_e^*$, $m_h^*$) such as indium arsenide (InAs) are promising[21]. With a very high mass ratio of $m_h^*/m_e^*$ ~20 in InAs, electrons obtain a majority of excess energy[22]. In addition, InAs has a strong absorption coefficient and large parts of the solar spectrum can be absorbed due to its small bandgap.[23] Recently, cooling of hot electrons in pure InAs NWs has been studied using time resolved photoemission electron microscopy with high spatial and temporal resolution. Among other results, strong cooling of hot electrons via electron-hole scattering has been observed[24]. This technique, however, is inherently unsuitable for probing hot carriers in core-shell NWs since this method relies on emission of electrons close to the surface[24]. Here, we overcome this limit and study various aspects of light harvesting and carrier cooling in core-shell NWs using optical femtosecond (fs) pump-probe experiments, exemplified for a case study of InAs-AlAsSb core-shell NWs. Transient dynamics measured at probe energies below the bandgap track the electron temperature via the carrier effective optical mass and elucidate the LO-phonon population. Probe energies above the bandgap provide direct information on the excess energy relaxation of hot electrons. Additionally, data taken at various excitation intensities is used to identify and compare cooling mechanisms of bare InAs NWs and core-shell NWs. Furthermore, we investigate the carrier population in the electronic states of the shell, which are located at higher energies than the band edge of the core.

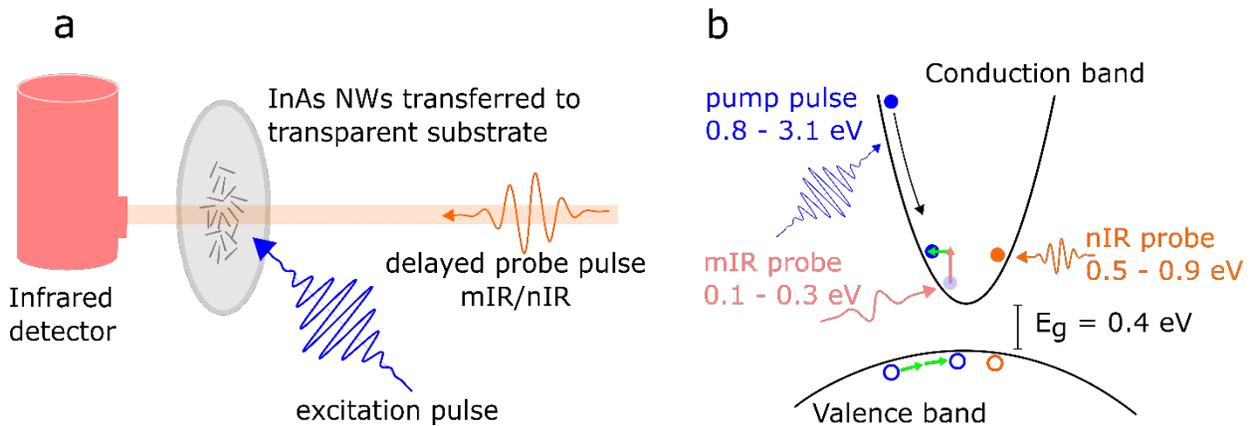

*Figure 1 a; Schematic of the transient absorption (TA) spectroscopy setup. A pump pulse excites hot carriers in InAs NWs which are randomly distributed on an infrared transmissive substrate. A delayed probe pulse passes the same spot and is detected. b; Pump and probe interactions as described in the band model. Probe energies below the bandgap (in mid-infrared (mIR) spectral range) are sensitive to free carrier absorption (FCA) while probe energies above the bandgap (in the near-infrared (nIR)) elucidate the conduction band occupation. Phonons (green arrows) are required for momentum conservation in the FCA process and are generated by the cooling of hot holes.*

## Experimental Design

Figure 1a depicts the transient absorption (TA) spectroscopy setup used in this work. A pump pulse (150 fs) in a broad range of energies (0.8-3.1 eV) delivered by a Ti:sapphire based laser system is used to excite the sample. Every second pump pulse is blocked by a chopper; thus, comparison of adjacent infrared pulses yields transient absorption of the photoexcited carriers. Positive TA values represent decreased transmission upon excitation while negative TA, i.e., bleaching, indicates increased transmission of the probe beam. Figure 1b shows excitation and probe mechanisms in a simplified band structure plot. The pump pulse excites an electron hole pair with excess energy, while two different probe beams are employed to resolve the carrier cooling and recombination dynamics on ultrafast timescales. The near infrared (nIR) probe beam has an energy (~0.5-0.9 eV) above the bandgap and is absorbed by interband transitions. Photoexcited carriers in the conduction band weaken this transition and cause a bleaching in nIR-TA, which reflects the energy distribution of electrons in CB. In contrast, the mIR-TA (~0.1-0.3 eV) has two major contributions: i) intraband free carrier absorption (FCA), which involves phonons (green arrow in Figure 1b) for momentum conservation, creates positive TA; ii) negative TA caused by the carrier induced reduction of the surface reflection. The former signal (i) tracks the phonon population, while (ii) monitors the carrier effective optical mass, a measure for the hot electron temperature. The temporal delay between pump and probe pulse is scanned up to 2 ns and transient absorption is measured at each delay. Obtained transient curves represent, thus, the dynamics of photoexcited carriers. Further details of the setup are described in the Methods section. All NWs explored in this study were transferred with very high areal density from their native growth substrates onto transparent $BaF_2$ substrates to capture the TA characteristics exclusively from the NWs. In total, we have investigated samples consisting of few-µm long InAs NWs with either ~70 nm and 120 nm diameter, as well as InAs-AlAsSb core-shell NWs with similar core diameters (40 nm, 140 nm), and an AlAsSb shell thickness of approximately 20-25 nm. In addition, a polished InAs wafer was used as bulk reference. All NW samples were grown with very high homogeneity in a completely catalyst-free process using selective area epitaxy (SAE)[25,26] (see section 1 in the SI), resulting in a wurtzite dominated crystal phase and a high density of stacking faults. The low-temperature wurtzite InAs bandgap energy estimated from previous photoluminescence (PL) experiments is approximately ~0.44-0.46 eV[26,27] for NWs grown under similar conditions.

**Transient Absorption below Bandgap**

First, as depicted in Figure 2a, mid-infrared transient absorption of bare InAs NWs has been measured for near resonant excitation with only 0.35 eV excess energy. The probe wavelength of 8 μm (150 meV) is far below the bandgap and, thus, only sensitive to FCA. One can observe a strong negative TA signal upon excitation which declines to zero within few ns. Decays are clearly carrier density dependent as larger excitation intensities cause a faster decline. Negative TA of the mIR probe is unexpected, since photoexcited carriers enhance FCA and reduce transmission. However, photoexcited carriers also lower the refractive index at the NW surface and therefore increase transmission via a decrease in reflection[28]. Transient absorption due to FCA increases with the probe wavelength and positive TA is observed for λ > 10 μm (for more details, see Figure S5 and section 4 in the SI). Passivated NWs with an AlAsSb shell show positive TA for all mIR probe wavelengths, since carriers are confined inside the core, which prevents a reduction of the refractive index at the surface, and thus exhibit weak free carrier absorption. Previous studies have measured TA of vertical standing NW arrays[29]; in this geometry the probe beam passes through the NW length which can be up to 50 times greater than the diameter. Therefore, FCA is much stronger than transient reflectivity on the surface and only positive TA has been observed[30].

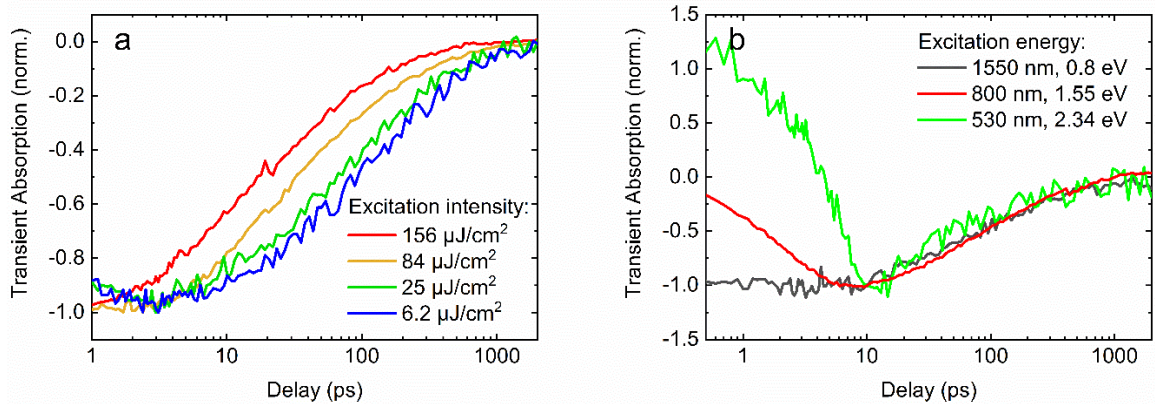

*Figure 2: a; Excitation intensity dependent TA of bare InAs NWs excited with 1550 nm pulses and probed at 8 μm. A strong negative TA signal (increased transmission) is observed upon excitation, all curves are normalized with respect to their absorption minimum. b; Excitation with excess energy results in a short-lived absorption superimposed with the long-lived negative signal. Data has been measured at a probe wavelength of 7 μm and was normalized to the negative TA signal at 10 ps.*

Recombination of photoexcited carriers leads to a decline of the negative TA signal. Charge carrier recombination in similar NWs has been investigated before[29], and can be analyzed using the common ABC-rate equation model involving Auger, bimolecular and trap-assisted recombination (see Figure S6, section 5 in the SI). Even without further analysis, the decreasing decay rate with decreasing excitation intensity clearly points towards relevant contributions by bimolecular and Auger-recombination. Auger recombination rates of InAs NWs determined by the rate equation model are consistent with previous studies on bare core-only NWs and are two orders of magnitude lower than Auger-rates of bulk InAs[29,31].

We can use the excitation intensity dependent recombination for a comparison of transient absorption curves obtained for different excitation wavelengths, as shown by the mIR TA data of Figure 2b. A positive TA signal emerges at short delays for small excitation wavelengths. The additional absorption vanishes within 10 ps, leaving only negative TA signal due to reduced surface reflection. Data has been normalized at 10 ps to highlight the similar decline of negative TA for all curves on the hundred-ps timescale. Although experiments with multiple excitation beams are usually hard to compare, due to varying spot sizes and absorption coefficients, the similar decline of the bleaching signal proves almost the same charge carrier densities for all three curves. Therefore, the short-lived additional absorption can be directly attributed to excitation with excess energy, or in other words: hot carriers. We propose phonon assisted free carrier absorption (PFCA) as a source of this signal[32]. In order to conserve momentum, the absorption of mid-infrared photons by electrons involves phonons. Thus, the measured absorption signal is bilinearly proportional to both populations i.e., of electrons and phonons[33]. Following excitation with excess energy, electrons and holes cool *via* emission of LO-phonons which in turn increase the cross section for free carrier absorption[34]. Owing to the absence of strong negative TA by charge carriers, core-shell NWs are, thus, better suited to investigate dynamics of the PFCA. Hereby, the temporal evolution can be described by a convolution of LO phonon generation with decay of the excess LO-phonon population *into* acoustic phonon modes[35]. Multiple observations indicate that the observed LO phonons are emitted by hot holes, instead of hot electrons: First, the phonon signal reaches its maximum fast, within 1 ps, which indicates that the driving cooling process occurs on the same timescale. It is already known that holes cool much faster than electrons due to the more densely packed states in the valence band[36]. Moreover, we observe cooling of electrons on slower timescales (see following sections). Secondly, the wavevector of phonons emitted by cooling of hot electrons is too small to satisfy momentum in the mIR-FCA process, but the wavevector of phonons emitted by holes is larger due to the smaller curvature of the VB. In our work, PFCA dynamics are measured for various excitation wavelengths and extrapolation to zero excess energy – neglecting ongoing phonon generation – yields a mean lifetime of 2.3 ps for LO phonons. The same value is obtained by exponential fits applied to the decline of the PFCA signal. For more details see Figures S7, S8 and the discussion presented in section 6 in the SI. Time-resolved Raman spectroscopy and Monte-Carlo simulations reported a comparable LO-phonon lifetime of 1.8 ps for bulk InAs[37]. As the phonon signal stems most likely from cooling of hot holes, we cannot make statements regarding the phonon population emitted by hot electrons, which is relevant for a hot phonon bottleneck. However, we assume that the lifetime of 2.3 ps for LO-phonons is valid for the entire LO-phonon population, independent of the phonon wavevector.

## Cooling of Hot Electrons tracked by Effective Mass

As mentioned before, negative mIR-TA is observed for unpassivated NWs because photoexcited carriers lower the refractive index and reflection at the NW surface. In this section, we will quantify this phenomenon and extract the temperature of the (hot) electron population. In addition, we compare InAs NWs, whose change in reflectance is determined in transmission, with a bulk InAs sample, since its polished surface allows direct transient reflection measurements. A first order expansion of Fresnel's formula shows, that the carrier-induced change in reflectivity ($\Delta R$) is almost linear to the change in refractive index ($\Delta n$)[38].

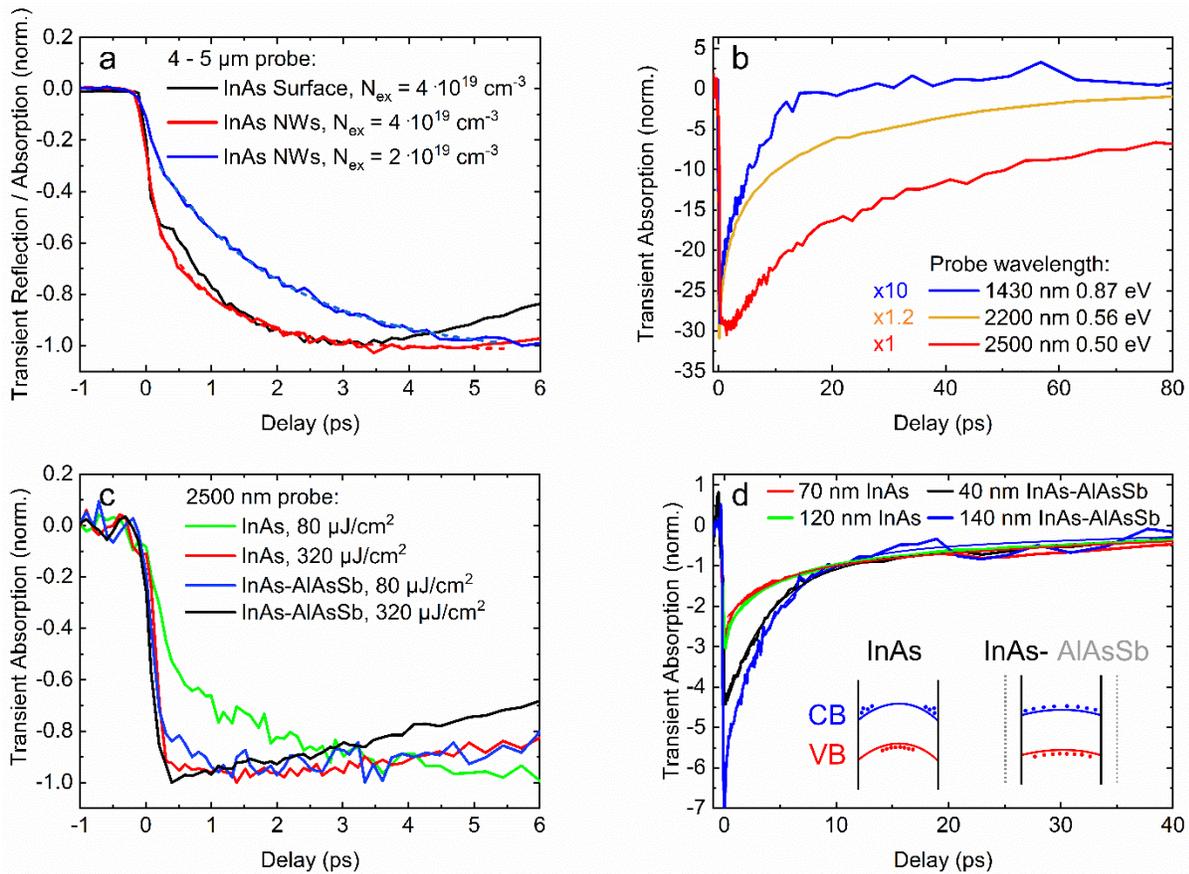

*Figure 3: Transient absorption measured after 800 nm excitation. a: mIR-TA of bare InAs NWs (diameter 120 nm) measured at two excitation intensities, generating different photoexcited carrier densities $N_{ex}$. mIR transient reflection data measured at a bulk InAs surface are shown in black for comparison. Dashed lines are exponential fits. b: Normalized bleaching curves obtained by*

*various nIR-probe wavelengths for 120 nm-thick InAs NWs excited at 800 nm (240 µJ/cm²) with corresponding scaling factors. c: Bleaching measured at a probe energy close to the bandgap for pure InAs and core-shell NWs of thickness 120 nm and 140 nm, respectively. d: Conduction band dynamics measured at 2.2 µm for all NW samples excited at 280 µJ/cm². All curves have been normalized at 10 ps to highlight the shared, slow decay channel. Dashed lines show biexponential fits.*

$$\frac{\Delta R}{R} = \frac{4}{(n_0^2-1)} \Delta n \quad (1)$$

For mIR photons with an angular frequency ω far from the band resonance, $\Delta n$ is well described by the Drude model[39]

$$\Delta n = -\frac{2\pi e^2}{n_0 m_{opt}^* \omega^2} \Delta N \quad (2)$$

$$\frac{1}{m_{opt}^*} = \frac{1}{m_e^*} + \frac{1}{m_h^*} \quad (3)$$

$n_0$ is the refractive index of the unperturbed sample, $m_{e,h}^*$ are the temperature dependent effective masses of electrons and holes, respectively. Equation 2 shows that the refractive index is lowered by additional charge carriers $\Delta N$ and the deviation is inversely proportional to the carrier effective optical mass $m_{opt}^*$. Due to band non-parabolicity, described by the parameter α (~1.4 eV⁻¹)[22], the effective mass increases with temperature $T$:[40–42]

$$m_e(T) = m_{e0}(1 + 5\alpha k_B T) \quad (4)$$

Therefore, we observe an initial bleach upon photoexcitation of hot carriers and a delayed rise when carriers cool, as the effective mass is reduced and the change in refractive index reaches a maximum[38]. This can be seen in Figure 3a for the bulk InAs surface (transient reflection at 5 µm) and InAs NWs (transient absorption) probed at 4 µm. Bulk InAs (black) was excited with the same intensity (~240 µJ/cm²) as NWs (in red), while the blue data shown for NWs refers to lower excitation intensity (~120 µJ/cm²). Photoexcited carrier densities were determined from the excitation density as demonstrated in Section 5 of the SI. For a probe wavelength of 4 µm, we can neglect free carrier absorption and differential transmission is dominated by a change in reflection[43]. One can see that transient reflection at the surface of a bulk sample shows almost the same initial dynamics as transient absorption of InAs NWs. Note that thermalization and momentum scattering of photoexcited electrons occurs much faster than our time resolution and is expected to be finished within tens of femtoseconds[44]. At timescales accessible in our experiments, we expect that hot electrons are distributed such that one can assign a temperature. Smaller excitation intensities and therefore smaller carrier densities show a prolonged risetime, form this we conclude that the electron effective mass and temperature changes at a smaller rate. For delays exceeding few ps, the absolute value of the bleaching signal no longer increases but declines due to recombination of carriers. This decay is faster in bulk InAs due to diffusion of photoexcited electrons opposite to the surface, which reduces the local carrier density.

Another mechanism which could explain a delayed rise in transient reflectance is intervalley scattering with hot electrons returning from the L and X valley to the Γ valley. While this effect is of considerable interest for hot carrier solar cells[45], we assume that this process is of little relevance for the interpretation of our data as it is expected to be carrier density independent[46,47], thus contradicting our observation. Also, the relative number of carriers stored in side-valleys is small. The same argumentation can be applied to impact ionization, a process where hot carriers lose excess energy by exciting additional electron hole pairs[21]. This would create additional carriers shortly after excitation and increase the transient signal but has a low efficiency and is also considered density independent[48]. Although both effects might take place, they are obscured by the change of the refractive index during cooling of hot electrons.

Our excitation intensity dependent measurements show that fast cooling within the first 5 ps is a carrier-carrier interaction dominated process, as cooling rates decrease with decreasing concentrations of electrons and holes. Furthermore, it is the only predominant cooling mechanism since the data can be well described by single exponential fits (see solid lines in Figure 3a). In contrast, cooling of hot electrons by emission of LO-phonons cannot explain the density dependent behavior in Figure 3a, as this should be independent of the carrier concentration[36]. If a phonon bottleneck emerges, cooling rates should even decrease for larger excitation densities. With that, only electron-electron and electron-hole scattering remain, with the former being unable to cool the thermalized electron population. Electron-hole scattering, however, can redistribute energy, especially due to the large mass ratio of electron and holes. Hereby, electrons get a large fraction of the exciting photons' energy, which makes scattering of hot electrons at cold holes a major loss mechanism[49]. Recent work employing time-resolved electron emission spectroscopy on InAs bare-core NWs also identified e-h scattering as the main cooling mechanism of hot electrons occurring on the few picoseconds timescale[24].

Furthermore, we can estimate the temperature of the electron gas *via* the ratio of bleach at 200 fs (change in refractive index by hot carriers at $T_{initial}$) versus maximal bleach at 5 ps (change in refractive index by considerably cooler carriers). This evaluation is pursued in detail in Section 7 of the SI, and yields a temperature of 2600 K for the hot electron ensemble shortly after its generation, which is in good accordance with previous studies.[24] From the exponential process obtained by fitting rising flanks in Figure 3a, we can construct a simple model for the cooling of hot electrons. At each scattering event with cross section $\sigma_{e-h}$, an electron transfers a certain amount of its energy η to a hole. This leads to exponential cooling of the electron temperature ($T_e$):

$$\frac{\partial T_e}{\partial t} = -\lambda T; \lambda = \frac{1}{\tau} = N_h \sigma_{e-h} \eta \quad (5)$$

This model is supported by the observation, that cooling times obtained from exponential fits double from 1 ps to 2 ps if the photoexcited carrier concentration is reduced from $4 \cdot 10^{19} cm^{-3}$ to $2 \cdot 10^{19} cm^{-3}$. Cooling of hot electrons is not necessarily finished within this cooling time, but the small change in the effective mass associated with the remaining cooling process is covered up by carrier recombination which causes a decline of the signal.

## Energy and Time Resolved Conduction Band Occupation

The hot electron dynamics in core-shell (and bare core) NWs can be monitored *via* nIR-TA measurements with a probe energy above the bandgap where the photoexcited carriers bleach the probe absorption by reducing the difference in population between valence and conduction band[50]. Probing above the bandgap allows us to track the hot electron population cooling to the CBM. Figure 3b shows that bleaching declines faster for higher probe energies, as states high above CBM are only occupied shortly after excitation. Probe energies closer than ~100 meV to the CBM (2500 nm probe wavelength) show slow dynamics which are governed by carrier recombination and band filling instead of cooling. Apart from a slower decay, probe wavelengths closer to the CBM also show a stronger change in absorption. This is because even at electron temperatures of several thousand Kelvin, the Fermi-Dirac occupation function strongly favors energetically low-lying states. Occupation probabilities are shown for several delays and probe wavelengths in Figure S10 in section 8 of the SI.

Figure 3c shows the rising flank of above bandgap bleach measured at 2500 nm probe wavelength, which is approximately 60 meV above the CBM. For NWs without shell, we observe a rise in absolute value when reducing the excitation intensity from 320 µJ/cm² to 80 µJ/cm². The same carrier density dependent effect takes place in core-shell NWs, but the rising flank seems to be faster overall. Electrons which are excited with almost 1 eV excess energy (800 nm excitation) require a few ps of relaxation until occupation near the band edge reaches its maximum. Therefore, the rising flank of bleaching near CBM is an indicator for cooling of hot electrons. Excitation intensity dependent dynamics, as already observed for the effective electron mass, indicate that initial cooling in unpassivated and core-shell NWs is based on a carrier-carrier interaction, namely electron-hole scattering. Interestingly, passivating NWs with a shell seems to accelerate this cooling mechanism – we will discuss this later along with the decay of above bandgap bleach.

Since lifetimes of hot electrons depend strongly on the selected probe wavelength, we have modeled the bleach dynamics with Fermi-Dirac occupation and obtained that a probe wavelength of 2.2 µm represents the temporal evolution of electron temperature best (see Figure S9a). Probing the band gap bleach at 1.4 µm yields lifetimes which are only a third of the electron temperature's lifetime as only the hot tail is observed. Vice versa, longer probe wavelengths yield longer lifetimes and represent partly recombination and band filling instead of cooling. Now we compare hot electron cooling in all NW samples by probing them at 2.2 µm; fits applied to these transients will directly describe the electron temperature (see Figure 3d). Two exponential functions are required to fit the data, which indicates another cooling mechanism on long timescales in addition to electron-hole scattering described in the sections before. Consistent with previous studies, we assign this to electron-phonon interactions[24]. Up to this point, electrons have thermalized, and electron-hole scattering diminishes the difference in temperature between these two carrier species. Hence, in the last step, electrons will thermalize with the lattice by phonon emission. This process is expected to be material specific and thus similar in all NW samples, as observed by the similar decline for delays greater 10 ps in Figure 3d.

In contrast, we observe that the initial cooling via e-h interactions within the first 10 ps is very sensitive to the NW-structure and is enhanced in core-shell NWs (see Figures 3c and 3d). Energy exchange between electrons and holes obviously depends on the carrier concentration and spatial overlap of both carrier populations. A less pronounced influence on initial electron cooling is given by the diameters of the NWs, with thinner ones cooling slower regardless of a shell. The data are analyzed *via* a biexponential fit to account for e-h interactions ($t_1$) and electron-phonon cooling ($t_2$). Our global fitting procedure yields $t_1 = (3.8 \pm 0.1)$ ps and $t_2 = (35 \pm 2)$ ps for e-h interactions and electron-phonon cooling, respectively. Since $t_1$ is carrier density dependent, deviations from literature values can be easily explained by different excitation conditions. In this case, however, our lifetime is in good agreement with previously reported values of 2.7 ps[24]. The cooling time obtained for electron-phonon cooling should be density independent, if we are below the threshold for a hot-phonon bottleneck, and our extracted values are in good agreement with previous optical studies on hot electrons in bulk InAs[32]. Time-resolved electron photoemission of InAs NWs yields only 14 ps for electron-phonon interactions, which is considerably shorter[24]. One explanation might be that both transient absorption above the band-edge and photoemission probe the high energy tail of the electron distribution, thus providing shorter decay times than the average thermalization time of the hot electron ensample. We propose that transient absorption measured at 160 meV above CBM, is more sensitive for cooling processes occurring on long timescales than time resolved photoemission, measured at 1.4 eV above CBM, since a much larger fraction of the electron population is monitored.

*Table 1: Ratio of amplitudes of fast versus slow decay channel obtained from biexponential fits for the investigated NW samples (60 nm and 120 nm refer to the diameters of the bare, unpassivated InAs NWs, while 40 nm and 120 nm with shell correspond to the diameters of the InAs core in passivated InAs-AlAsSb core-shell NWs).*

| Sample | 70 nm | 120 nm | 40 nm with shell | 140 nm with shell |
|---|---|---|---|---|
| $A_{e-h}/A_{Phonon}$ | 0.88 | 1.25 | 3.2 | 5.66 |

The fit amplitudes displayed in Table 1 clearly show the increased contribution of electron-hole scattering in the cooling process in core-shell NWs. Hereby, the strong influence of the passivation indicates a clear surface mediated effect. To some extent, shell induced changes in reflectivity could create higher initial carrier densities in passivated NWs which cause larger electron-hole scattering rates. Since the refractive index is comparable for core and shell, we propose that this effect is negligible[51]. Instead, we take a closer look at the core-shell interface. It is well known that InAs surfaces contain high n-type surface defect densities, which pin the surface Fermi level above the CBM, leading to downward bending of the conduction and valence bands[52]. This Fermi-level pinning is particularly important for InAs NWs, given their large surface to volume ratio, as has been recently assessed quantitatively. [52] Therefore, intrinsic InAs NWs have n-type character as the conduction band at the surface is below the Fermi-level[53]. Further, holes accumulate at the center of NWs which leads to a spatial separation of holes and electrons[23]. This effect is illustrated by a sketch in Figure 3d. We propose that this effectively reduces e-h scattering in unpassivated

InAs NWs by reducing electron hole wavefunction overlap and scattering rates. On the other hand, when InAs NWs are passivated by a shell layer, almost flat band conditions are obtained in the core, and carriers in the valence and conduction band are less spatially separated with reduced screening of electron-hole scattering. Hence, our observations of reduced e-h scattering in unpassivated NWs, along with the general findings that radiative electron-hole recombination is strongly enhanced by surface passivation, supports the notion of spatial separation of electrons and holes due to Fermi-level pinning[52]. Similar suppression of cooling *via* e-h interactions has been previously reported for quantum dots with hole selective trapping[54,55] and negatively charged nanocrystals with hole deficit[56], but has never been associated with Fermi-level pinning due to the different material nature of the investigated nanostructures.

The same explanation can be used to describe the size dependency of the fast cooling process which roughly scales with the NW diameter for both unpassivated and core-shell NWs (see Table 1). Instead of changing the band structure at the surface, as in case of the shell, thin wires possess a larger surface to volume ratio, thus, more hot electrons, excited across the NW volume, can accumulate at the surface where electron-hole scattering is maximal screened. Another explanation might be, that quantization effects start to play a role in the limit of thin NWs by reducing e-h scattering, as predicted by previous theoretical investigations.[9] Yet, these quantum confinement effects are still relatively weak for the range of the even thinnest NWs investigated here[57], and further transient absorption studies into ultrathin InAs NWs are required to provide additional proof.

## Charge Carrier Dynamics of Shell Associated States

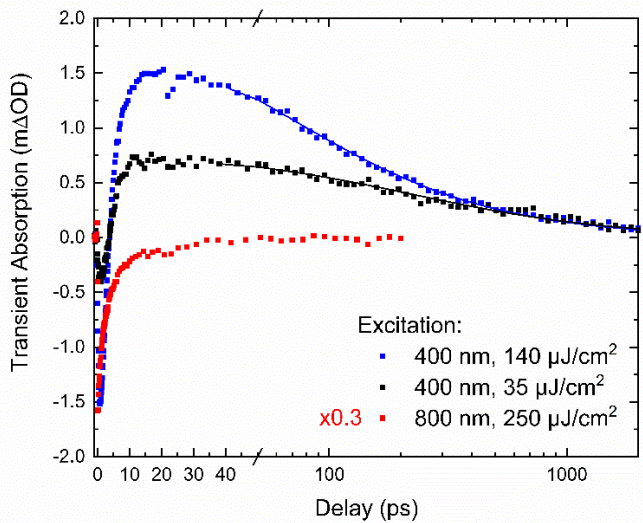

*Figure 4: Excitation of core-shell NWs (140 nm core diameter) at 400 nm results in an additional, long lived transient absorption signal compared to excitation at 800 nm. The decline is well described by bimolecular recombination (fit lines).*

While mainly investigating hot carrier dynamics associated with the InAs NW core, we found an additional population of charge carriers exclusively in core-shell NWs. As depicted by the excitation energy dependent data in Figure 4, we observe evidence of charge carriers within the shell for high-energy excitation at 400 nm, with a probe energy at 2 μm.

Note, that for the arsenic content (16%) of our lattice-matched AlAsSb shell, the indirect bandgap is at approximately 1.7 eV and the direct bandgap at 2.2 eV[58]. This leads to a large conduction band offset of ~1.2 eV, whereas the valence band maxima between core and shell are expected to be aligned. The long-lived TA signal is only observed if two conditions are fulfilled: i; excitation at 530 nm or a shorter wavelength and ii; the presence of a shell in the sample. This suggests that with short excitation wavelengths (e.g., 400 nm) carriers are excited within the AlAsSb shell, and hence associated shell states can be probed. Here, the probe wavelength neither equals electronic transitions nor free carrier absorption but matches a surface plasmon resonance of these NWs (see Figure S4a, Section 3 in the SI). Photoexcited carriers in the shell alter the carrier density and, thereby, influence the surface plasmon strength and position. The short-lived bleach in Figure 4 observed at delays smaller than 10 ps has its origin in the conduction band occupation of the core. Interestingly, the long-lived positive TA signal (up to several 100 ps) observed in Fig.4 suggests that electrons generated at an energy above the bandgap of the shell material do not migrate into the core but remain inside the shell although the shell CB is almost 1.2 eV above the CBM of the core. Further, we conclude a low trap density from the strict bimolecular recombination dynamic of carriers located in shell states (see fit lines).

It should be noted that the carrier lifetime in shell states was on the order of hundred ps in our experiments, but due to the density dependence minority carrier lifetimes could be well in the tens of nanosecond regime, which may allow extraction in solar cells. These observations highlight the possibility of tandem cells with multiple absorber materials within a single nanowire. Efficient light absorption by shell states could be realized *via* tailored plasmon modes[59,60]. Indeed, strong surface plasmon modes were observed in UV-Vis extinction spectroscopy for all NW samples, as illustrated in Figure S3 in the SI. Hereby, the spectral position of the resonant absorption feature is only determined by the NW diameter and the refractive index of the environment. Absorption spectra (Figure S3) and a more detailed discussion are presented in section 3 in the SI. By aligning surface plasmon resonances with shell associated band transitions, light absorption of shell layers could be greatly improved.

**Conclusions**

We employed femtosecond pump-probe spectroscopy in the visible-mIR spectral region to observe cooling of hot carriers in differential absorption measurements of InAs and InAs-AlAsSb core-shell NWs. Cooling of photoexcited, hot holes is most likely indirectly observed *via* the fast buildup of an excess phonon population. This phonon population can be tracked *via* phonon assisted free carrier absorption and allows us to determine the LO-phonon lifetime, on the order of ~2.3 ps. Cooling of hot electrons is observed *via* the evolution of the effective optical mass and occupation of high-lying states in the conduction band. We find a carrier density dependent fast cooling channel (~ 3 ps) which is assigned to energy transfer from electrons towards holes and a slower decay channel which is assigned to electron-phonon interactions (~ 30 ps). The interpretation of cooling channels and obtained lifetimes are in good agreement with reported values in literature[24], whereas our method allows to extend observation of cooling dynamics to core-shell NWs. We find increased contribution of e-h interactions for core-shell NWs and similar cooling rates via phonon emission, as compared to unpassivated InAs NWs. We attribute the change in electron-hole interactions to their spatial separation induced by Fermi-level pinning in bare, unpassivated NWs. Apart from carrier cooling, we found that core-shell NWs can absorb photon energies way above the core bandgap with shell states preserving most of the excess energy. Shell associated states show a pure bimolecular recombination which excludes a high trap density and should lead to long lifetimes for ambient carrier densities. In addition, UV-VIS extinction spectroscopy shows strong surface plasmon modes for all NWs studied. These can be spectrally tuned by varying the thickness and tailored such that they match inter-band transitions of the shell which allows efficient light absorption by thin shell layers. Finally, we want to emphasize that although e-h scattering is a major loss mechanism for excess energy of hot electrons in our experiments, this is unlikely to occur in most applications as carrier densities are usually much lower. Optical pump probe experiments of all kinds excite very dense carrier populations which are by far more affected by carrier-carrier interactions than steady-state systems. In identifying the electron-hole interaction as cooling pathway, this will open research efforts to resolve whether this effect will be as pronounced under ambient conditions and to what extent could limit hot carrier applications at low carrier densities.

**Methods**

InAs NWs with thicknesses between 40 – 140 nm and lengths up to 8 µm were grown by selective-area molecular beam epitaxy (MBE) on silicon[25]. Selected samples were further passivated by a lattice matched $AlAs_{0.18}Sb_{0.82}$ shell (20 nm) and capped with a thin layer of GaSb (5 nm) in order to prevent oxidation[26]. Exemplary scanning electron microscopy (SEM) images and further details about selected growth conditions are depicted in Section 1 in the SI. Previous transmission electron microscopy performed on these NWs reveals characteristic stacking of dominant wurtzite phase interrupted by a high density of stacking defects[25,26]. For reference measurements, a bulk n-type InAs epi-wafer was also used.

As-grown NWs were mechanically transferred to barium fluoride ($BaF_2$) windows to allow pump-probe experiments in a wide range of excitation energy without dealing with photoexcited carriers in the silicon substrate. Additionally, transferring NWs avoids Fabry Perot effects in optical spectroscopy by the otherwise periodic arrangement of as-grown NW arrays on freestanding substrate[61]. SEM measurements were performed to confirm that the NWs were not destructed during transfer.

Femtosecond transient absorption was measured employing a Ti:sapphire CPA laser system with 1 kHz repetition rate and 150 fs pulse length (Coherent Libra). Using various non-linear effects, a broad range of pump wavelengths (400 nm – 1550 nm) and probe wavelengths (either 1.4 µm – 2.5 µm by optical parametric amplification (OPA) or 3.7 µm – 12 µm by OPA) and successive difference frequency generation (DFG) was generated. Infrared probe light was dispersed by a grating spectrometer and detected by a liquid nitrogen cooled MCT array. Time resolution was achieved by varying the beampath of the pump beam relative to the probe beam with a delay stage up to 2 ns. Pump and probe beam were focused on the sample and overlapped at the focus. 90 % of the pump beam's intensity was within a diameter of 200 - 300 µm depending on the excitation wavelength. The excitation intensity was varied using a half-wave plate and a polarizer. A chopper at half repetition rate was used to block every second pump pulse, and transient absorption was calculated by comparing probe intensities with and without pump.

## Supporting Information

Additional information includes microscopic characterization of NWs, absorption measurements, a rate model for carrier recombination, extrapolation of LO phonon lifetimes, estimation of hot electron temperatures and derivations for cooling times from above bandgap bleaching.

## Acknowledgements

This research was supported by the Deutsche Forschungsgemeinschaft (DFG) via the Cluster of Excellence "e-conversion" (EXC 2089/1-390776260), the TUM International Graduate School of Science and Engineering (IGSSE) via project 11.05 (NW-Solar) and the Institute for Advanced Study (TUM-IAS) via the Focal Period 2019 program. M.N. thanks the "Studienstiftung des deutschen Volkes" for a PhD scholarship. This project has received funding from the European Union's Horizon 2020 research and innovation programme under the Marie Skłodowska-Curie grant agreement No 899987.

Supporting Information for:

# Hot Carrier Dynamics in InAs-AlAsSb Core-Shell Nanowires

*Daniel Sandner[1], Hamidreza Esmaielpour[2], Fabio del Giudice[2], Matthias Nuber[1], Reinhard Kienberger[1], Gregor Koblmüller[2] and Hristo Iglev[1,*]*

[1] Chair for Laser and X-ray Physics, Physics Department, Technical University of Munich,

James-Franck-Str. 1, 85748 Garching, Germany

[2] Walter Schottky Institute and Physics Department, Technical University of Munich, Am

Coloumbwall 4, 85748 Garching, Germany

- Corresponding authors: E-Mail: hristo.iglev@tum.de; gregor.koblmueller@wsi.tum.de

# Table of Contents

## Section 1: Microscopic characterization of as-grown NWs

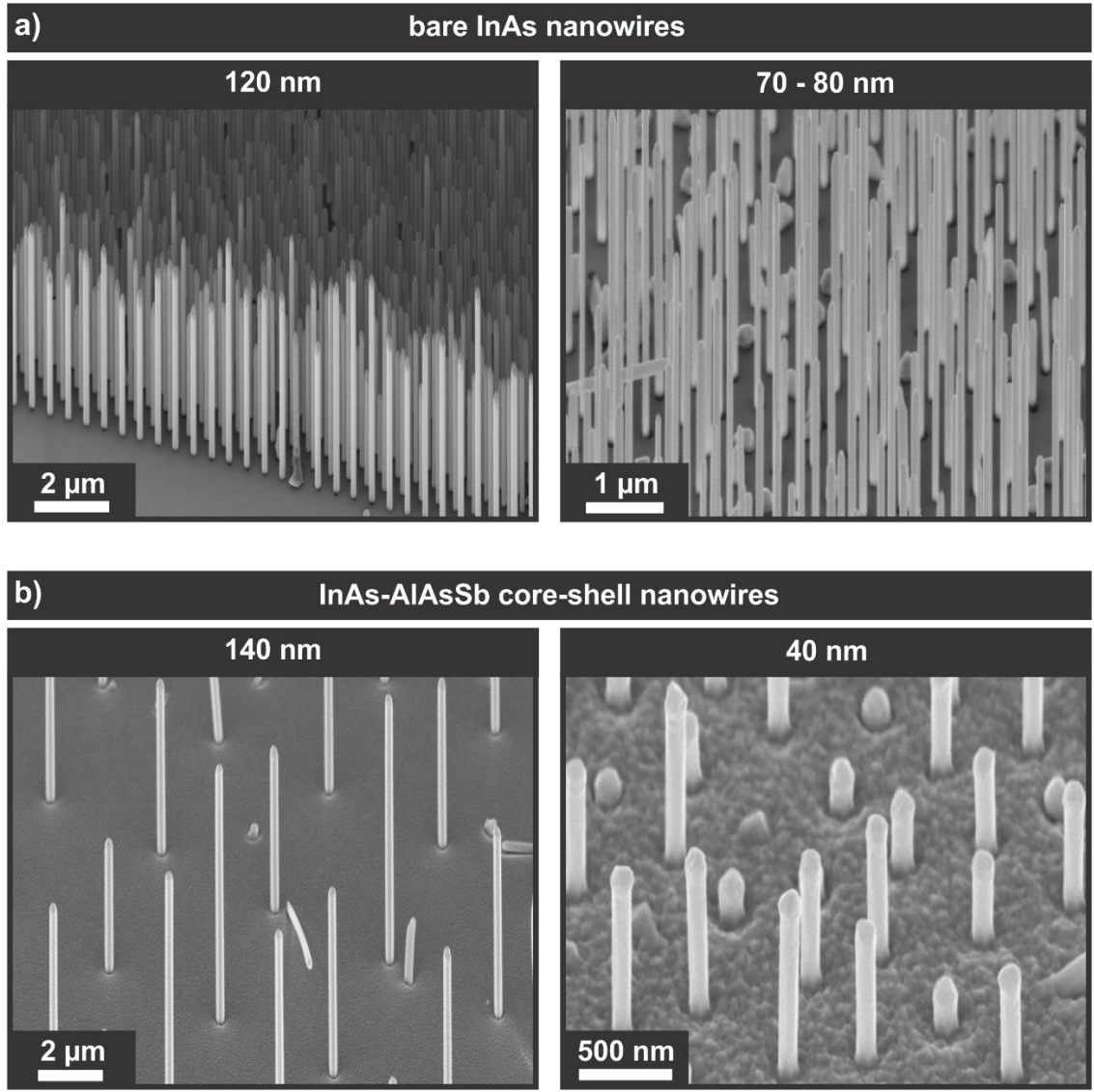

*Figure S1: SEM images of as-grown NW arrays on silicon. (a) shows bare, unpassivated InAs NWs with average diameters of ~120 nm (left) and ~70 nm (right), respectively, while (b) illustrates InAs-AlAsSb core-shell NWs containing NW cores with diameters of ~140 nm (left) and ~40 nm (right). The latter are overall shorter due to the reduced growth time necessary to realize very thin cores.*

All NWs were grown as periodic arrays on Si (111) substrate using selective-area molecular beam epitaxy (SA-MBE)[1]. To realize site-selective growth, the Si (111) substrates were prepatterned by nanoimprint lithography (NIL) or electron beam lithography (EBL) to create mask patterns with different pitches. Growth of the InAs NWs was then performed at a temperature of 520°C, In flux of 0.6 A/s, and $As_4$-BEP (beam equivalent pressure) of 4.5e-5 mbar. Scanning electron microscopy (SEM) images of two bare (core-only) InAs NWs are shown in Fig. S1(a) after a growth duration of 60 min. Average NW diameters and lengths for the two respective samples are ~120-nm and ~4-5 µm (left), and ~70 nm and ~3-4 µm (right), respectively.

Regarding InAs-AlAsSb core-shell NWs, the NW cores are grown under the same conditions. One sample had identical core growth time of 60 min, resulting in ~140-nm thick and ~7-8 µm long InAs NW cores (Fig.S1(b), left), while for another sample the growth time was only 7.5 min to achieve very thin NW cores of ~40-nm diameter (Fig.S1(b), right). Due to the reduced growth time, the NWs are overall much shorter (<1 µm). The cores in both samples were overgrown for 30 min with an AlAsSb shell at growth temperature of 425 °C and for another 10 min with a GaSb cap layer to prevent oxidation of the AlAsSb. The Al and Ga fluxes were set at 0.5 A/s, and the $Sb_2$-BEP was 1.2e-6 mbar. The As cell was set to conditions that yield an incorporated As-molar fraction of ~18% in the AlAsSb shell, which represents close lattice matching conditions with the InAs core[2]. Under these growth conditions, the thicknesses of the AlAsSb (GaSb) shell (cap) layers grown in the radial direction are on the order of ~20-25 nm (~5-nm). Hence, the overall diameters of these core-shell NW samples appear by about ~50-60 nm thicker than their respective NW cores.

## Section 2: Infrared absorption spectra

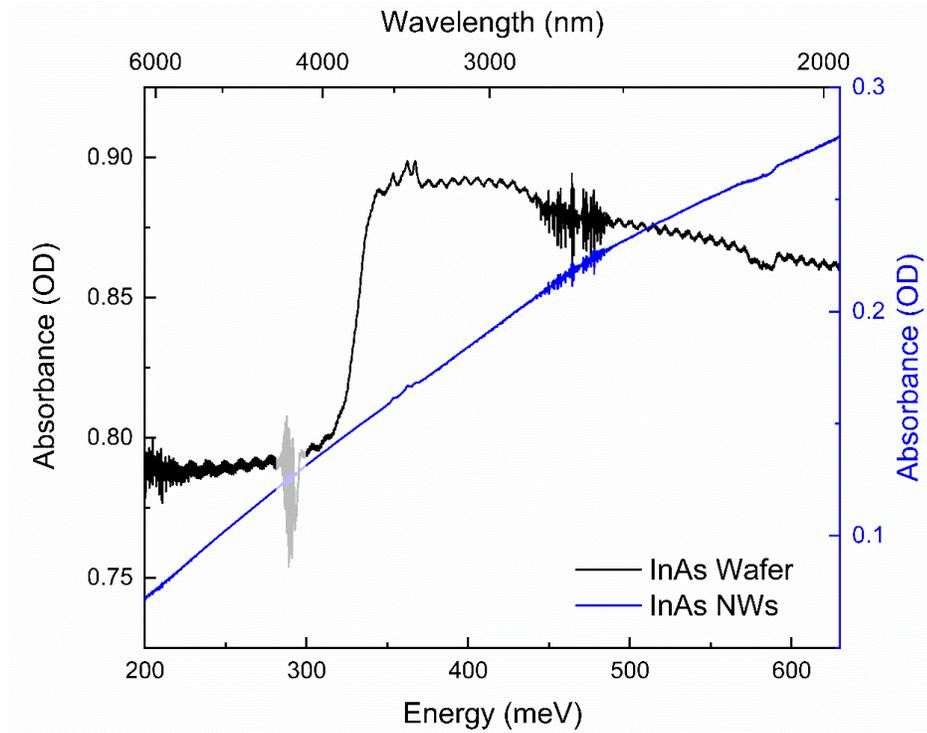

*Figure S2: Infrared absorption spectra measured for bulk InAs epi-wafer and unpassivated InAs NWs of diameter 120 nm. A $CO_2$ absorption artefact at 280 meV is faded.*

Fourier transform infrared (FTIR) spectra were measured for all NW samples and a bulk InAs wafer as reference at room-temperature. Figure S2 shows the onset of band absorption at 330 meV in the bulk sample, which corresponds to the literature value for the InAs bandgap at room temperature[3]. The spectra of all NW samples lack the distinct band edge. Instead, the absorption increases with the light frequency due to scattering.

## Section 3: Visible optical absorption (Surface Plasmon Resonance)

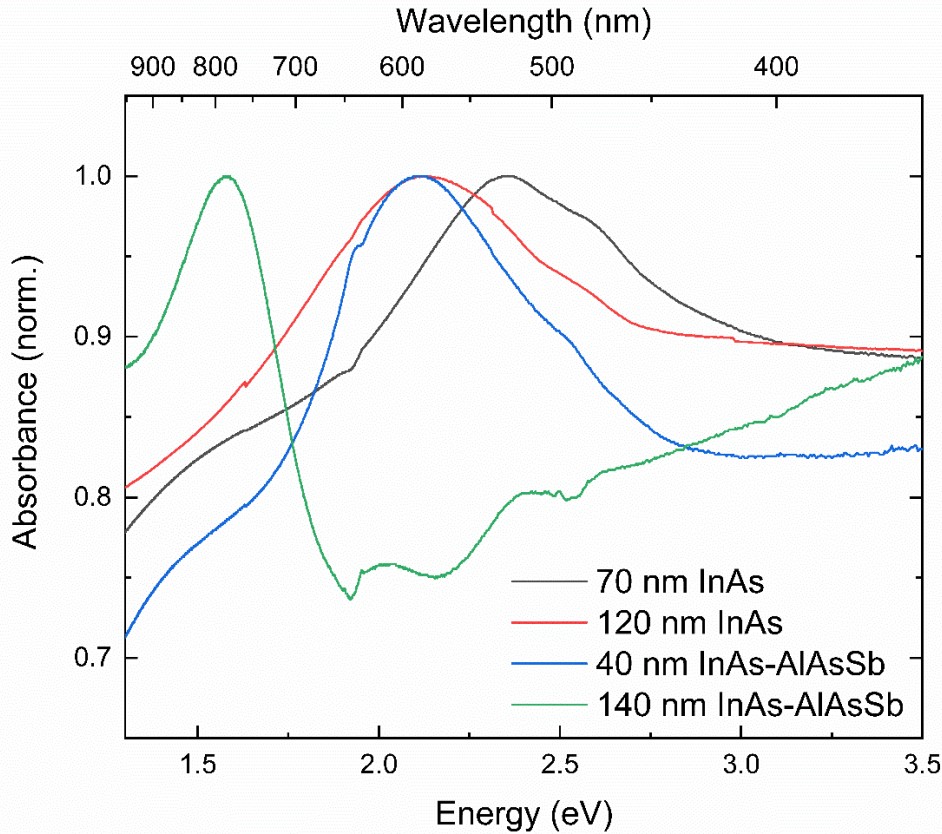

*Figure S3: UV-Vis absorbance spectra of NWs of several size after transfer to transparent Barium fluoride (BaF$_2$) windows.*

Figure S3 shows absorbance spectra of NWs transferred to BaF$_2$ windows as measured by UV-Vis extinction spectroscopy. One observes a symmetrical peak for each sample. The spectral position of the resonance depends on the thickness, surface passivation, and does not coincide with an interband transition (~0.4 eV for InAs). Two trends are visible: first, thin NWs show a blueshifted resonance compared to thick ones, under the same set of surface conditions. Secondly, in the presence of a shell, the resonance is redshifted. This is explained by the altered refractive index induced by surface passivation[4]. Both observations can be interpreted to stem from surface plasmon resonances (SPR). In addition, passivation narrows the spectral width of the observed peaks as the defect density at the surface is reduced and plasmons are less damped. Surface plasmon resonances are very sensitive to the molecular environment and may extend the potential of InAs NWs for sensing applications. Previous InAs NW based sensors used FET-like structures and recorded changes in carrier mobility associated to chemical bonding [5]. Surface plasmons may allow the use of NW arrays as sensors with optical read out. Furthermore, the absorption of surface plasmons can exceed those of thin shell layers and can be tailored for efficient light coupling.

Optical microscopy, presented in Figure S4a, shows that following mechanical transfer of NWs, most of the 140 nm-thick NWs with a shell kept a certain orientation. This is also reflected in UV-Vis and FTIR spectroscopy performed with linear polarized light, seen in Fig. S4b. One can see that certain absorption features are independent of the NW orientation (e.g. at 1.7 eV), while others (0.6 eV, 1.2 eV, 2.5 eV) depend clearly on the polarization of light relative to the NW axis. We attribute the absorption features to various surface plasmon modes. The strong influence of the thickness on the SPR is supported by the observation that the SPR absorbance is stronger for light polarized perpendicular to the NW axis.

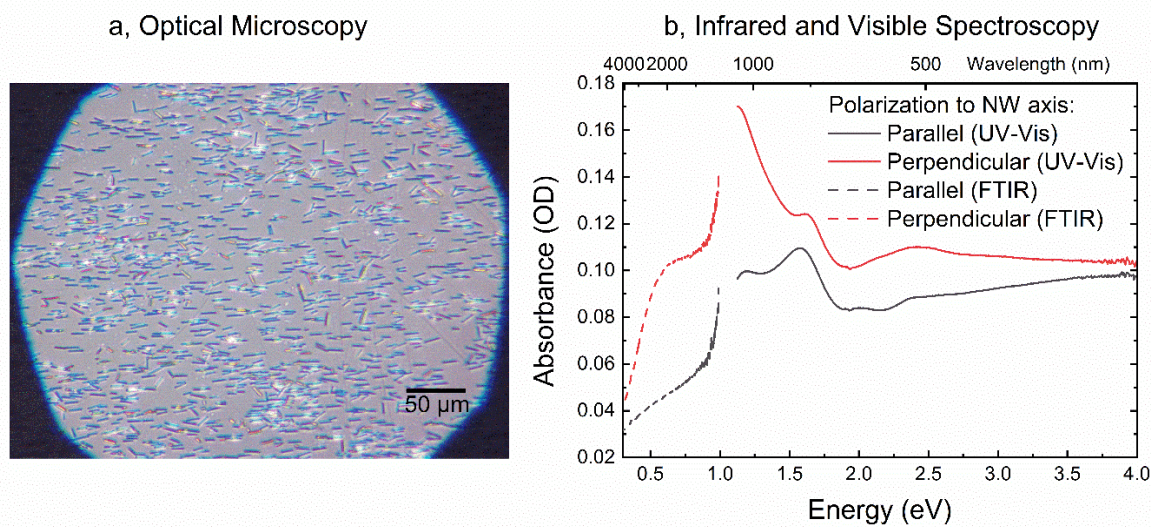

*Figure S4 a; Optical microscopy of 140 nm-thick NWs with shell after transfer to a barium fluoride substrate. Most NWs are orientated in the same direction. b; UV-Vis extinction measured with linear polarized light parallel and perpendicular to the majority NW axis.*

## Section 4: Transient absorption signatures of photoexcited carriers:

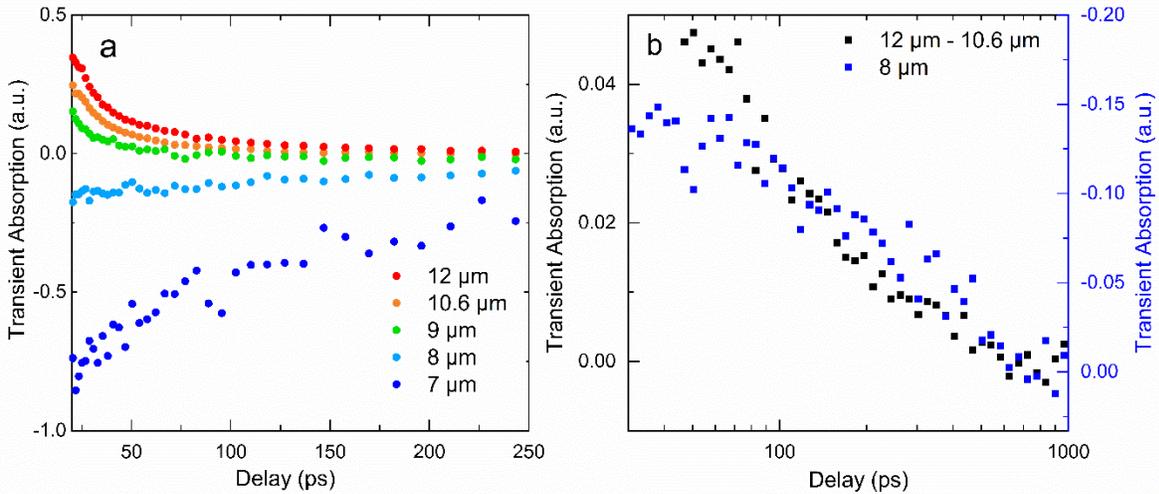

*Figure S5: a; Depending on the probe wavelength, photoexcited carriers in InAs NWs cause increased absorption or transmission of a mid-infrared probe beam. InAs NWs with a thickness of 120 nm were excited at 400 nm with an intensity of ~200 µJ/cm². b; Photoinduced bleaching (blue, right y-axis) and absorption (black, left side) show the same decay dynamics, namely carrier recombination.*

In mIR transient absorption spectroscopy, two concurring effects arise from photoexcited carriers in InAs NWs. On the one hand, the refractive index is lowered for mIR wavelengths, this reduces reflection at the surface and increases transmission. On the other hand, free carrier absorption of the carriers reduces transmission. Which effect is dominant, depends on the probe wavelength, as FCA is wavelength dependent.

Figure S5a shows transient absorption of bare core, 120 nm-thick NWs, excited at 400 nm with an intensity of ~200 µJ/cm2, for probe wavelengths in the range 7 - 12 µm. One can see that negative transient absorption due to reduced reflectivity is dominant for $\lambda < 9$ µm. For probe wavelengths larger than 10 µm, FCA causes a positive TA. In Figure S5b, bleaching dynamics are plotted on the right y-axis (note that the scale is flipped) and absorption (as difference between 12 µm and 10.6 µm) is plotted on the left y-axis. Only delays exceeding 25 ps are shown, since excess energy by 400 nm excitation causes hot carrier effects (PFCA) on early delays. One can observe, that despite the flipped sign, all probe wavelengths monitor the same decline dynamics. Decline of the photoexcited signal in transient absorption is in most cases attributed to carrier recombination, thus both bleaching and absorption following excitation are sensitive to additional charge carriers generated by the pump pulse.

Some materials, which exhibit temperature dependent absorbance, show an additional signal upon photoexcitation due to deposited energy (heat). Then, decline of the signal represents actual cooling of the pump spot by heat diffusion. To exclude this effect, temperature dependent FTIR spectra of the NW samples were measured in the range 290 – 350 K, showing no temperature induced absorption.

## Section 5: Carrier recombination and ABC-model

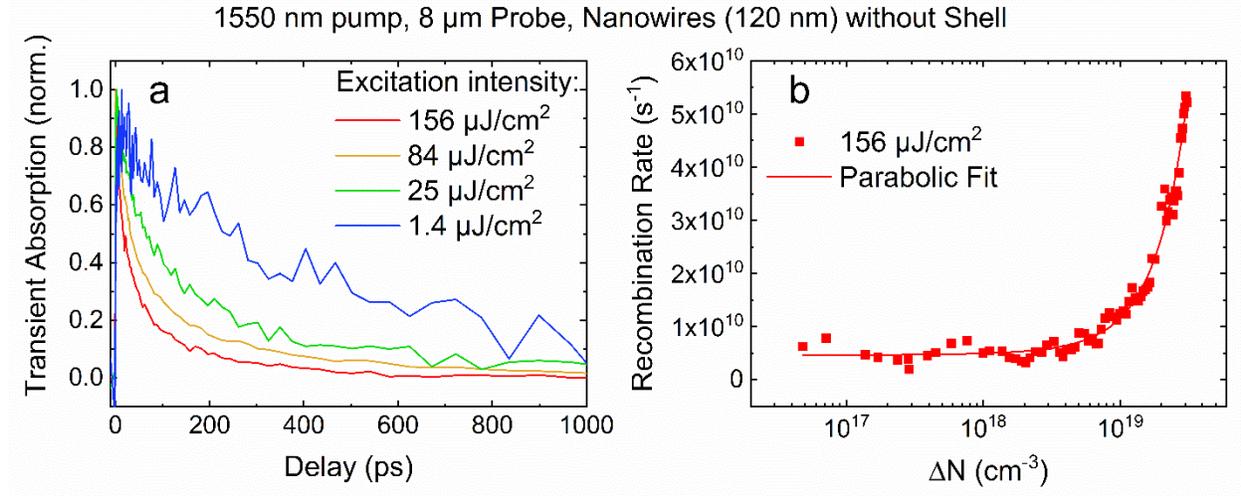

*Figure S6: a; Transient absorption of bare core 120 nm-thick NWs measured after 1550 nm (0.8 eV) excitation with 8 µm (150 meV) probe for various excitation intensities. Curves are normalized regarding the maximum. b; Recombination rate, calculated from transient absorption measured at 156 µJ/cm², plotted versus carrier density, a parabolic fit is used to determine different recombination contributions.*

Recombination of charge carriers was studied using near-infrared excitation, thus, photoexcited carriers have little excess energy. For a probe wavelength of 8 µm, the carrier induced change in reflectivity causes negative TA, but curves were normalized to the range (0,1). Figure S6a shows carrier density dependent behavior of the recombination dynamics of unpassivated 120-nm thick InAs NWs, where lower excitation intensities clearly show increased lifetimes. This indicates carrier-carrier interactions as recombination paths, namely bimolecular and Auger recombination, in addition to trap assisted Shockley-Read-Hall recombination.

To describe the recombination more quantitatively, we used the commonly employed ABC model[6]:

$$R = -\frac{\partial N / \partial t}{N} = A_{SRH} + B \cdot N + C \cdot N^2$$

$A_{SRH}$: Trap assisted recombination according to Shockley-Read-Hall mechanism, non-radiative recombination *via* mid-gap defect states. Since most defects are near the surface, it is useful to compare the surface recombination velocity S: $S = \frac{A_{SRH} d}{4}$, which eliminates the size dependency of $A_{SRH}$ for NWs

B: Bimolecular (radiative) recombination coefficient, C: Auger recombination coefficient

We exemplify this for the highest investigated excitation intensity (156 µJ/cm²), where the transient absorption curve has been smoothed and derived with respect to the delay for the total recombination rate R (see Figure S6b). A parabolic fit to the data yields the three respective coefficients, $A_{SRH}$, B, and C.

The carrier density is estimated *via* the absorption coefficient determined by photothermal deflection. This method determines the absorbance of a sample by its rise in temperature upon illumination. Small temperature changes are measured by the deflection of a laser beam passing the sample, immersed in perfluorohexane. The obtained absorption coefficient of $4 \cdot 10^4$ cm$^{-1}$ at 1500 nm is in good agreement with the absorption coefficient of bulk InAs in the literature[3]. For NWs with a thickness of 120 nm, we obtain that 60% of the pump pulse is absorbed, thus, the excitation intensity of 156 µJ/cm$^2$ generates additional carriers with a density of approximately $4 \cdot 10^{19}$ cm$^{-3}$. A previous study has shown that the change in refractive index and therefore bleaching scales almost linear with the excess carrier density[7]. Therefore we assume a linear relation between the observed bleaching and the photoexcited carrier density. Since the strongest change in absorption for excitation with 156 µJ/cm$^2$ was 10 mOD, we used $4 \cdot 10^{18}$ cm$^{-3}$/mOD as conversion factor for all other data points of the TA curve. For all experiments in this study, we estimated the photoexcited carrier density by the absorption coefficient and excitation density (calculated from pump power and beam size). It is not possible to use the carrier density/bleaching signal conversion factor for other experiments since the observed bleaching is the product of the charge carrier density and the number of NWs within the probe spot and the concentration of NWs across the sample is rather inhomogeneous due to the mechanical transfer.

*Table S1: Fit results of the rate model and comparison to literature:*

| $A_{SRH}$ (s$^{-1}$) | B (cm$^3$ s$^{-1}$) | C (cm$^6$ s$^{-1}$) | Reference |
|---|---|---|---|
| $(4.6 \pm 0.5) \cdot 10^9$ | $(3 \pm 1) \cdot 10^{-10}$ | $(4 \pm 0.3) \cdot 10^{-29}$ | this study |
| $(1.4 \pm 0.5) \cdot 10^{10}$ | $(1.42 \pm 0.16) \cdot 10^{-11}$ | $(8.23 \pm 0.44) \cdot 10^{-28}$ | InAs NW[6] |
|  | $1.1 \cdot 10^{-10}$ | $2.2 \cdot 10^{-27}$ | InAs Bulk[8] |

Our results according to the ABC model are in the same range as previously reported values. $A_{SRH}$ is trap associated and therefore depends on the NW diameter (surface defects) as well as surface passivation. The corresponding surface recombination velocity for 120 nm thick bare core NWs is 138 m/s.

While B is larger in our study, C is lower but both parameters have a strong interdependence in fitting. In the literature, another method was used to determine the bimolecular parameter B to avoid interdependence of B and C[6]. In summary, our experiments agree with previously reported carrier recombination in InAs NWs, for B and C we find differing values due to the interdependence in fitting and the lack of additional experimental data for B.

## Section 6: Phonon lifetime extracted from PFCA in core-shell NWs

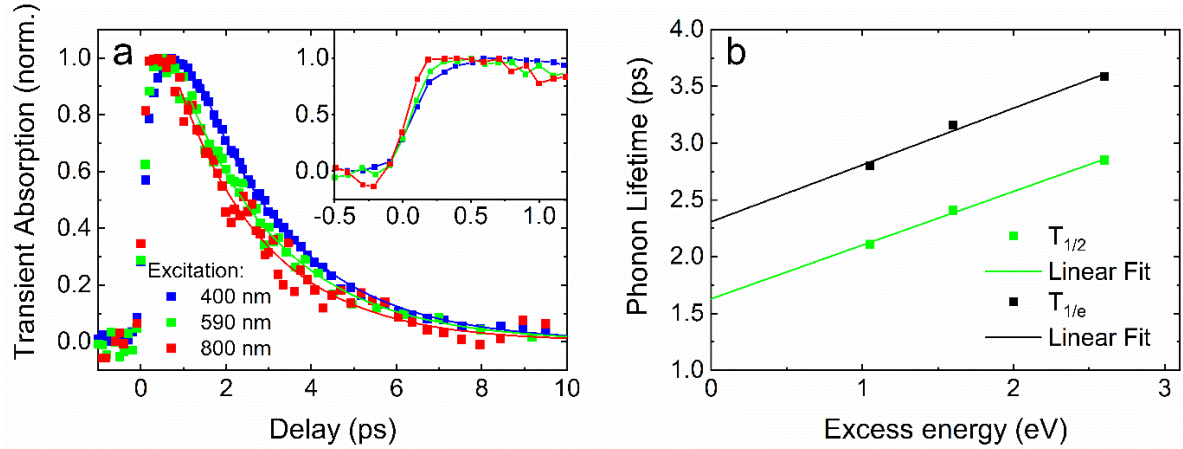

*Figure S7: a; Phonon assisted free carrier absorption measured in 40 nm thick core-shell NWs for several excitation wavelengths and a probe wavelength of 6 µm. The excitation intensity was ~600 µJ/cm² for 400 nm and ~300 µJ/cm² for 530/800 nm to account for the different photon energies. Curves are normalized to the maximum; solid lines are calculated by monoexponential fits. The inset shows the same data expanded at early delays. b; Mean lifetimes of the normalized curves, delays where the signal has dropped to ½ and 1/e, are plotted versus the excess energy ($E_{exc} - E_{gap}$). A linear fit extrapolates to zero excess energy.*

Core-shell NWs showed no negative TA in our experiments since the refractive index at the surface (shell) is not altered by photoexcited carriers inside the core. We observe solely weak free carrier absorption (FCA).

FCA is amplified by a nonthermal LO phonon population as momentum conservation requires phonons to participate in the FCA process[9]. As carriers cool, an excess of LO-phonons is generated which enhances the FCA cross-section of free carriers[10]. This phonon assisted FCA has been measured for three excitation wavelengths. Figure S7a shows that the decline of TA has the same shape for all three excitation wavelengths but starts later in time with increasing excess energy (see inset). Mono-exponential fits yield a lifetime of 2.3 ps for the decay. Decline is retarded at larger excitation energies due to the convolution of phonon decay with phonon generation, or in other words, the maximum is reached at later delays. Delays where the PFCA signal drops to 1/2 and 1/e have been plotted in Figure S7b versus the excess energy (excitation energy with subtracted bandgap). Extrapolation of excess energy to zero (excess phonons only once produced at t = 0) yields the same lifetime of 2.3 ps. The 1/e data intercepts the y-axis at 2.3 ps, and the 1/2 data at 1.6 ps. Both values indicate a lifetime of 2.3 ps for LO phonons ($e^{-1.6/2.3} = 0.5$), which agrees with previous literature reports. For example, Raman measurements of the LO-phonon mode of InAs NWs obtained a full-width at half maximum (FWHM) of approximately 4 cm⁻¹.[11] This equals a dephasing time of approximately 2.6 ps ($T_2 = \frac{1}{\pi c \delta_v}$). In literature there is only one reference which extracts a LO-phonon lifetime of 1.8 ps in bulk InAs *via* pump probe experiments and ensemble Monte Carlo calculations[12].

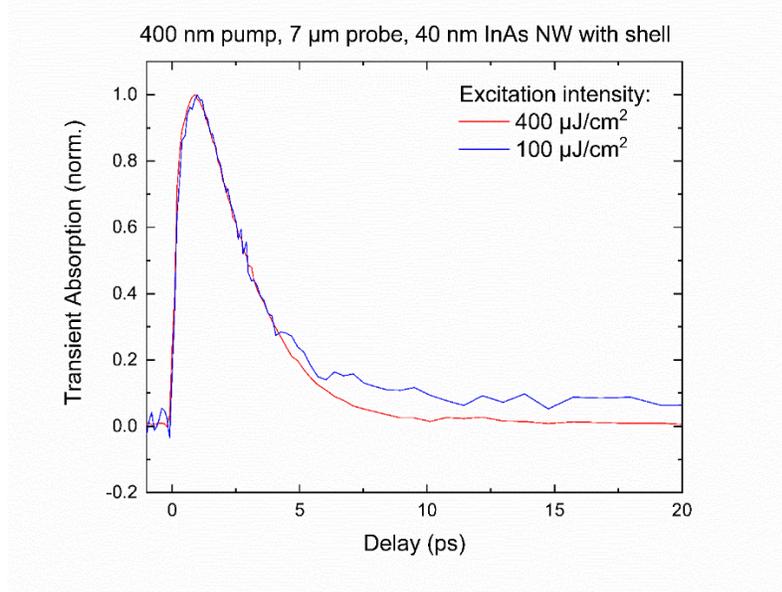

*Figure S8: Excitation intensity dependent transient absorption curves of 40 nm-thick core-shell NWs measured after excitation at 400 nm with 7 µm probe wavelength. Curves were normalized to the maximum.*

Excitation intensity dependent experiments (see Figure S8), show no distinct power dependency for the PFCA signal. Both curves recorded at very different intensities (400 µJ/cm$^2$, 100 µJ/cm$^2$) were normalized to their maximum. We assume that the long-lived tail (ordinary FCA) is more pronounced for the 50 µW data because PFCA does not scale exactly linear. Owing to normalization to the PFCA maximum, FCA appears stronger. Still, the temporal dynamics of phonon generation and decline up to 5 ps are indistinguishable for both excitation intensities.

Hot phonon bottlenecks have been observed so far in GaAs, but rarely in InAs[13,14]. We propose that the shorter lifetime of LO-phonons (2.3 ps for InAs, ~5-7 ps for GaAs[14]) requires larger carrier densities or more excess energy to establish and maintain a a hot phonon bottleneck.

So far, we have only discussed the decline of the observed PFCA signal. The generation of out of equilibrium phonons requires energy, which must be delivered by another non-thermal species. Following photoexcitation, hot electrons and holes are present in the sample. It would be natural to associate the generated phonons to cooling of hot electrons as they obtain most of the excess energy. However, there are several arguments against this. First, the phonon signal has a fast-rising flank and reaches its maximum within 1 ps. This would require hot electrons to transfer their energy very fast towards LO-phonons and would indicate that electrons are rather cold after 1 ps, as no more notable phonon generation takes place, and we observe only decay of phonons. This contrasts with our findings of significantly hot electrons on timescales well exceeding few ps and the observation, that the shortest cooling mechanisms of hot electrons is carrier-density dependent (see evolution of effective electron mass in the main text) and increases at larger carrier densities. LO-phonon emission instead should be carrier density independent or even decrease at large densities in the presence of a hot phonon bottleneck. Furthermore, LO-phonons emitted by hot electrons have a small wavevector which corresponds to the vertical shift of around 30 meV (LO-phonon energy) and the large slope of the conduction band. This wavevector is too small to account for the momentum needed in the FCA process, where a photon delivers a vertical shift of around 150 meV and the slope of the CB is at least the same, or even smaller, requiring more momentum. This is a strong hint that hot electrons are not responsible for the generation of LO-phonons which we observe in PFCA. Hot holes, however, remedy all issues described above. It is well known that cooling of hot holes occurs fast via phonon emission by the more densely packed states in the VB[15]. This would explain the fast buildup of the phonon signal. Moreover, the VB has a smaller curvature compared to CB and thus, hot holes emit LO-phonons with a larger wavevector, which could participate in the PFCA process.

If hot holes act as driving mechanism of the phonon generation, we can exploit the slope of the straight line in Figure S7b for cooling rates of hot holes. This figure shows the delayed decay depending on excess energy and the slope indicates a regression of 2 eV/ps. Note, that the excess energy was determined by subtracting the bandgap from the excitation energy. We need to consider that only approx. 20% of this excess energy is transferred to the hole due to the difference in effective mass between electrons and holes. The cooling rate of hot holes would then be 0.4 eV/ps. By the phonon energy of ~30 meV, we obtain a mean hole-phonon scattering time of ~75 fs.

## Section 7: Estimation of carrier temperature *via* effective electron mass

Using band non-parabolicity α, one can derive the temperature dependent optical mass of electrons [16] The dispersion relation in the CB reads:

$$E(1 + \alpha E) = \frac{\hbar^2 k^2}{2m_{e0}}$$

The electron effective mass is given by $m_e^* = \hbar^2/(\partial E^2/\partial k^2)$:

$$m_e(E) = m_{e0}\left(1 + \frac{10}{3}\alpha E\right)$$

$$m_e(T) = m_{e0}(1 + 5\alpha k_B T)$$

In the last step, E was replaced by $3/2\, k_B T$.

With previously derived linear relationships ($\Delta R \propto \Delta n, \Delta n \propto 1/m_{opt}^*$), it is possible to extract an initial electron temperature from the ratio of initial to total bleach:

$$\frac{\Delta n_{in}}{\Delta n_{final}} = \frac{1/m_e^*(T_{in})}{1/m_e^*(T_{final})} = \frac{1 + 5\alpha k_B T_{final}}{1 + 5\alpha k_B T_{in}}$$

This calculation yields 2600 K for the delayed rise at 800 nm excitation and 1000 K for excitation at 1550 nm. By comparing excess energy in excitation and initial electron temperatures, we find a constant ratio of approximately 2500 K/eV.

# Section 8: Electron cooling time constants from above bandgap occupation

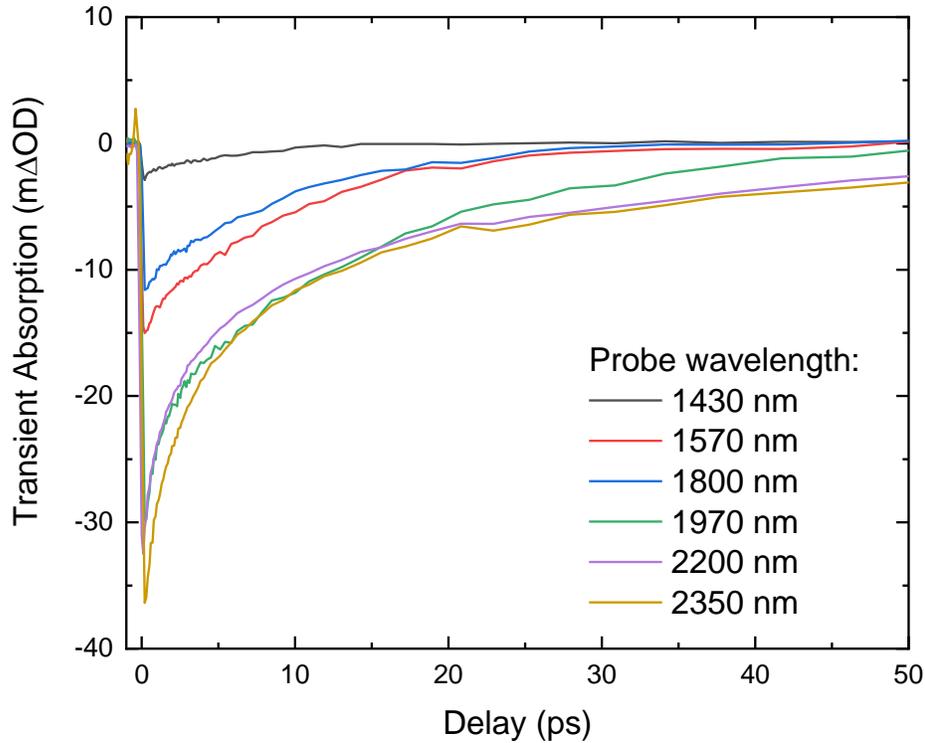

*Figure S9: nIR TA dynamics of 120 nm thick InAs NWs measured after excitation at 800 nm for several probe wavelengths. Selected curves are displayed in Figure 3 b.*

Figure S9 shows transient absorption measured for probe wavelengths above the bandgap (nIR). Excitation at 800 nm generates hot electrons which populate the conduction band. This reduces the difference in population between VB and CB compared to a sample which has not been excited. In consequence, the band absorption experienced by the probe beam, is weakened (bleached) for the excited sample and we observe increased transmission. States high above CBM are only occupied for a short period of time and, as electron cool, transient absorption measured for large probe energies (short wavelengths) declines fast towards zero. Further, states probed at a large probe energy have only a small probability for occupation which creates smaller absolute TA signals compared to probe energies close to CBM.

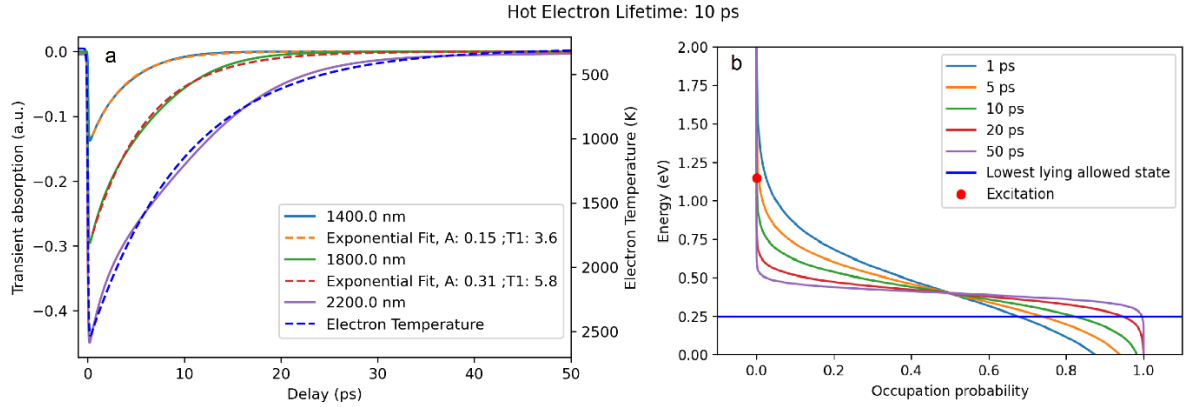

*Figure S10a: Simulated band absorption bleaching curves (as observed with the nIR probe) on the left y-axis and decline of electron temperature in the model on the right y-axis. Bleaching curves for 1400 nm and 1800 nm were fitted with a single exponential. b; Occupation probability of states depending on their energy, measured regarding the valence band maximum, for certain delays. The same function (exponential decay) for the electron temperature was used as in S9a. Excitation with 800 nm and the lowest allowed states of the conduction band, by surface band bending, are marked.*

Since decay times observed in transient near-infrared absorption strongly depend on the probe wavelength, we used Fermi-Dirac statistics to find the probe wavelength which represents electron temperature dynamics best. For this simple model, we assume that the electron temperature is 2600 K at t= 0 and decays with a single exponential function of 10 ps lifetime to 300 K. We then calculate the occupation probability according to Fermi-Dirac statistics for consecutive time steps and transient absorption for several probe wavelengths based on the diminished difference in population between CB and VB. Bandgap renormalization effects are not included.

Compared to our experiments, simulated transient absorption displayed in Fig S10a follows the same trends, namely lifetimes and amplitudes increase for larger probe wavelengths (see Figure S9). Simulated transient absorption curves were fitted with single exponentials, this reveals that transient absorption measured at a probe wavelength of 1400 nm underestimates the lifetime of hot electrons by a threefold (3.6 ps decay time of TA compared to 10 ps for electron temperature). Transient absorption modeled at 2200 nm, however, follows the electron temperature quite well. Fig S10b shows occupation probabilities for several delays depending on the energy of the state, measured relative to the VBM. Thermalization is not shown as it requires a much more advanced model but is expected to be finished well within 100 fs. One can see that the electron distribution is increasingly narrowed to the CBM due to cooling. The excitation at 800 nm (yielding the initial temperature of 2600 K in our experiments) and the lowest lying state of the CB (below the bandgap due to surface band bending) are marked. Quantitative values for the down bending were extracted from[17].